\documentclass{article}

\usepackage[letterpaper,top=2cm,bottom=2cm,left=3cm,right=3cm,marginparwidth=2cm]{geometry}

\usepackage[T1]{fontenc}
\usepackage{mathpazo} 

\usepackage{amsmath}
\usepackage{amssymb}
\usepackage{mathtools}
\usepackage{physics}
\usepackage{bm} 
\usepackage{setspace} 
\usepackage{authblk} 
\usepackage{chemformula}
\usepackage[version=4,arrows=pgf-filled, textfontname=sffamily, mathfontname=mathsf]{mhchem}

\usepackage{array} 
\usepackage{booktabs} 
\usepackage{caption} 
\usepackage{subcaption} 
\usepackage{float} 
\usepackage[section]{placeins} 

\usepackage{graphicx}
\usepackage{xcolor}
\definecolor{red}{rgb}{0.9,0,0}
\definecolor{magenta}{rgb}{1.0,0,1.0}

\usepackage{hyperref}
\hypersetup{
    colorlinks=true,
    citecolor=cyan,
    linkcolor=blue,
    urlcolor=magenta
}
\usepackage{cleveref}

\usepackage{enumitem} 
\usepackage{todonotes} 
\usepackage{changes} 
\usepackage{braket} 
\usepackage{tcolorbox} 

\usepackage[sorting=none, citestyle=numeric-comp]{biblatex}
\usepackage{quantikz}

\newlist{todolist}{itemize}{2}
\setlist[todolist]{label=$\square$}
\usepackage{authblk}

\title{Assessing Finite Scalability in Early Fault-Tolerant Quantum Computing for Homogeneous Catalysts}

\author[1]{Yanbing Zhou}
\affil[1]{Zapata Computing, Inc., 100 Federal St., Boston, MA 02110, USA}
\author[1]{Athena Caesura}
\author[2]{Corneliu Buda}
\author[2]{Xavier Jackson}
\affil[2]{Innovation and Digital Science, bp Technology, 501 Westlake Park Blvd, Houston, TX 77079, USA}
\author[2]{Clena M. Abuan}
\author[2]{Shangjie Guo\textsuperscript{*}}
\date{}

\begin{document} 

\maketitle
\begin{abstract}
As quantum hardware advances toward fault-tolerant operation, an intermediate stage known as early fault-tolerant quantum computing (EFTQC) is emerging, where partial error correction enables meaningful computation. 
In this regime, the ability of quantum processors to scale in size and depth has become a crucial factor shaping their achievable performance.
This study investigates how finite scalability influences resource requirements for simulating open-shell catalytic systems using Quantum Phase Estimation (QPE). 
The analysis compares hardware archetypes distinguished by fidelity or operation speed under two representative scalability models.
Finite scalability increases qubit and runtime demands yet leaves overall scaling behavior intact, with high-fidelity architectures requiring lower minimum scalability to solve equally sized problems.
These effects are largely independent of the chosen scalability model. 
Extending this framework, we examine runtime competitiveness across hardware and code configurations, incorporating surface-code and quantum Low-Density Parity-Check (LDPC)-based fault tolerance under finite scalability.
The results identify operating regimes where high-fidelity architectures remain competitive despite slower gate speeds and show that LDPC codes further expand this regime by reducing space-time overhead.
Together, these findings highlight the central role of scalability in quantifying performance and guiding the design of next-generation quantum hardware. Continued progress in scalable architectures will be essential for extending quantum computing to increasingly complex scientific and industrial applications.

\end{abstract}
\vspace{\fill}
\pagebreak

\section{Introduction}
The evolution of quantum computing from the Noisy Intermediate-Scale Quantum (NISQ) era toward fully fault-tolerant operation has revealed a complex landscape of technological advances and constraints. While NISQ processors enable proof-of-principle demonstrations, their limited coherence and lack of error correction preclude scalable quantum advantage. The long-term objective of Fault-Tolerant Quantum Computing (FTQC) is to overcome these limitations through large-scale, error-corrected computations capable of addressing practical problems in science and industry. Between these two regimes lies an increasingly important intermediate stage known as Early Fault-Tolerant Quantum Computing (EFTQC).

EFTQC marks a transitional phase in which partial error correction allows meaningful algorithms to be executed before full-scale fault tolerance becomes achievable. Recent studies~\cite{EFTQC_2021,EFTQC_2022,EFTQC_2024} have demonstrated that devices comprising tens of thousands to millions of physical qubits can already perform useful computations despite incomplete error correction, highlighting EFTQC as a critical bridge between experimental prototypes and fully protected architectures.

A central challenge in this transition is scalability, which governs how the accuracy and efficiency of quantum computations evolve as system size increases. In this context, scalability describes how the performance of quantum operations changes when more qubits and gates are added to a processor, reflecting the interplay between physical noise, control precision, and architectural design. 
In practice, scale-dependent effects such as frequency crowding in superconducting qubits and cross-mode coupling in trapped-ion systems can degrade gate fidelity and coherence as devices grow larger~\cite{Fellous-Asiani2021}.
These limitations determine how well quantum processors can sustain low error rates and deep circuits, making scalability a decisive factor in whether early fault-tolerant systems can achieve meaningful computational advantage.
Quantitative models of scalability~\cite{katabarwa2023} have recently been proposed to link hardware performance with algorithmic resource demands, particularly for algorithms such as Quantum Phase Estimation (QPE)~\cite{cleve1998quantum}. These models provide a framework for analyzing how scaling assumptions influence the reach and efficiency of quantum algorithms, offering guidance for optimizing resource allocation and selecting suitable hardware architectures in the transition toward fully fault-tolerant quantum computing.

In this work, we investigate how finite scalability affects the practical deployment of fault-tolerant algorithms for solving industrially relevant problems, with a particular focus on modeling open-shell transition-metal catalysts. 
We focus on how different scalability models affect quantum resource estimation and how these effects vary across two representative hardware archetypes: \emph{type A}, high-fidelity but slower devices, and \emph{type B}, faster but lower-fidelity architectures. 
By applying both finite power-law and finite logarithmic scalability models to \emph{type A} and \emph{type B} quantum hardware architectures, we evaluate the robustness of the scalability framework and its effectiveness in guiding hardware selection for complex quantum chemistry tasks.

Beyond resource estimation, we extend our analysis to the comparative performance of different error-correcting code architectures and hardware platforms. By modeling surface codes and modern LDPC codes within the scalability framework, we quantify how code efficiency and fault-tolerant overhead reshape the balance between devices with different fidelity and speed. This analysis identifies the operating regimes where \emph{type A} systems, despite slower gates, can remain runtime-competitive through superior error suppression, and where improved code efficiency further shifts this balance. In doing so, we connect scalability, code design, and architectural performance as jointly determining factors in the achievable capabilities of early fault-tolerant quantum computers.

Our objectives are threefold:
\begin{itemize}
    \item Assessing the robustness of the scalability model for different quantum architectures.
    \item Examining the impact of incorporating scalability into quantum resource estimation.
    \item Analyzing how scalability, architectural characteristics, and error-correcting code choices jointly influence hardware performance and algorithmic feasibility.
\end{itemize}

By situating this work within the broader transition from NISQ to FTQC, we highlight scalability as a defining factor that links hardware performance with long-term architectural strategy. As systems scale, maintaining control over error accumulation becomes as critical as increasing qubit counts, since the effectiveness of scalability ultimately depends on how well hardware performance and error mitigation coevolve. This relationship determines whether emerging EFTQC platforms can support complex quantum chemistry applications with both precision and efficiency.

\section{Technical Background}
This section introduces the key concepts required to understand the application of finite scalability to quantum resource estimation in the context of open-shell catalysis. 
We begin by outlining the surface code and quantum Low-Density Parity-Check (LDPC) codes and their roles in determining circuit depth through error correction. The theoretical foundations of the scalability model are then presented, followed by an introduction to the open-shell catalytic systems used as representative examples of industrially relevant catalysts.

\subsection{The Surface Code}
The surface code is among the most mature and practical approaches to reducing gate error and enabling large-scale quantum computation~\cite{fowler2012surface, horsman2012surface}. In quantum error correction, additional qubits redundantly encode quantum information, allowing errors to be detected and corrected during computation. The strength of a code is quantified by its distance $d$: if fewer than $d$ errors occur, they can be corrected. Each \emph{logical qubit} in the surface code requires $2d^2$ physical qubits, and the logical error rate decreases exponentially with $d$~\cite{fowler2012surface}.

Given quantum hardware with physical gate (or measurement) error rates of $p_{\text{phys}}$, the surface code can facilitate computation with logical gate (or measurement) error rates:
\begin{align}
\label{eq:logical_error_rate}
p_{\text{L}} &= 0.1 \left( \frac{p_\text{phys}}{p_{\textup{th}}} \right)^{\frac{d+1}{2}},
\end{align}
where $p_\text{th}$ is the surface code threshold, here taken as 0.01~\cite{goings2022, fowler2019}.
The probability of error-less computation $p_{\text{tot}}$ with $m$ gates using physical qubits is 
\begin{align}
    p_{\text{tot}} = 1 - (1 - p_{\text{phys}})^m \leq m p_{\text{phys}},
\end{align}
where the upper bound follows from the union bound, which tends to be extremely tight in the regime of error rates relevant to large-scale quantum computations.
If logical qubits is instead used, an exponentially favorable trade-off in the  circuit depth ($m$) versus the number of physical qubits from ~\cref{eq:logical_error_rate} can be obtained:

\begin{align}
\label{eq:exponential_suppression}
    m \propto \left(\frac{p_{\text{th}}}{p_\text{phys}}\right)^{\sqrt{\frac{n}{8k}}}.
\end{align}
However, this tradeoff is only available if $p_{\text{phys}} < p_{\text{th}}$ (i.e. below the~\emph{threshold} of the surface code).

In this work, we consider surface code gates implemented through lattice surgery operations~\cite{horsman2012surface, litinski2018lattice, litinski2019game}, chosen for their favorable overhead~\cite{fowler2018low} and well-established quantum resource estimation frameworks~\cite{litinski2019game}. Other two-qubit gate paradigms exist~\cite{fowler2012surface}, but lattice surgery provides a more scalable approach, as further discussed in Section~\ref{subsec:two_qubit_gates}.

\subsection{LDPC Codes}
\label{subsec:LDPC}

Quantum Low-Density Parity-Check (LDPC) codes offer higher thresholds and lower qubit overhead than surface codes~\cite{Babar2015, breuckmann2021quantum, bravyi2024high}. While surface codes rely on local four-qubit connectivity, LDPC codes require each qubit to maintain a small, constant number of long-range connections. 
This additional constraint on the connectivity of the device means that implementing LDPC codes on \emph{type A} devices such as ion traps is more likely to be achieved in the early fault-tolerant regime. 
In this work, we focus on Bivariate Bicycle (BB) LDPC codes for which Bravyi et al.~\cite{bravyi2024high} recently presented error-rate scaling relations.

The logical error rate for BB LDPC codes can be modeled as~\cite{bravyi2024high}
\begin{align}
\label{eqn:LDPC_logical_error_rate}
    p_L = \left(p_{\text{phys}}\right)^{\frac{d_{\text{circ}}}{2}} \cdot \exp\left(c_0 + c_1 \cdot p_{\text{phys}} + c_2 \cdot p_{\text{phys}}^2 \right)
\end{align}
where $d_\text{circ}$ is the circuit distance and $c_0$, $c_1$, and $c_2$ are fit parameters.
This model does not allow dynamic adjustment of the code distance; instead, we select from the five codes identified in Ref.~\cite{bravyi2024high}, choosing the smallest capable of executing the circuit.

When a device supports the long-range connectivity required for LDPC codes, we assume that these couplings can be exploited to implement transversal gates. Details of this assumption and its implications for quantum resource estimation are further discussed in Section~\ref{subsec:two_qubit_gates}.

\subsection{Two-Qubit Gate Implementations}
\label{subsec:two_qubit_gates}
\begin{figure}
    \centering
    \begin{quantikz}[column sep=0.5cm, row sep=0.5cm]
    \lstick{$\ket{\psi_1}$} & \ctrl{1} & \rstick{} \\
    \lstick{$\ket{\psi_2}$} & \targ{}  & \rstick{} 
    \end{quantikz}
    \quad = \quad
    \begin{quantikz}[column sep=0.5cm, row sep=0.5cm]
    \lstick[3]{$\ket{\psi_1}$}   & \ctrl{3} & \rstick{} & & \\
     &  & \ctrl{3} & \rstick{}  & \\
     & & & \ctrl{3} & \rstick{} \\
    \lstick[3]{$\ket{\psi_2}$}  & \targ{} & \rstick{} & & \\
     &  & \targ{} & \rstick{} & \\
     &  & & \targ{} & \rstick{}
    \end{quantikz}
    \caption{Schematic of a transversal two-qubit gate between logical qubits. The left panel shows a logical CNOT gate acting between $\ket{\psi_1}$ and $\ket{\psi_2}$, while the right panel illustrates its transversal implementation across the corresponding physical qubits that comprise each logical qubit.}
    \label{fig:transversal_gates}
\end{figure}
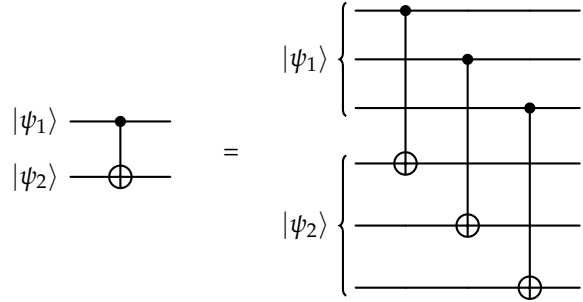

When quantum error-correcting codes were first conceived, two-qubit gates between logical qubits were envisioned to be implemented transversally, meaning that operations were performed directly between corresponding physical qubits. An example of this approach is shown in Fig.~\ref{fig:transversal_gates}.

In practice, most quantum devices are constrained by local connectivity, which limits the feasibility of fully transversal operations. To address this, lattice surgery~\cite{horsman2012surface} was introduced as a method for performing logical two-qubit operations using only local interactions. This technique involves measuring qubits along the shared boundary of two surface-code patches in order to ``merge'' and ``split'' logical qubits. Although the full details of lattice surgery are beyond the scope of this paper, its implications can be summarized in two key aspects.

First, lattice surgery introduces an overhead equal to the code distance $d$, arising from the additional stabilizer checks required during merging or splitting operations. Second, lattice-surgery-based operations are naturally described in the ZX-calculus framework rather than in standard circuit representations~\cite{de2020zx}. The first effect is straightforward to account for: it simply introduces a multiplicative factor of $d$ in the execution time.

Establishing a fair comparison between transversal and lattice-surgery-based implementations is more subtle. For simplicity, we assume a ``best case'' ion-trap architecture, where transversal gates can be performed freely between any pair of qubits. In this limit, the total number of two-qubit gates is approximately equal to the number of Clifford operations that must be executed\footnote{Idle steps are counted as Clifford operations, since the identity is a Clifford gate.}. This assumption effectively removes the ``routing overhead'' typically included in resource estimates for lattice surgery.

For both architectures, we further assume that the cost of magic-state distillation, in terms of the number of Clifford operations, is comparable. Although this assumption is imperfect, accounting for the cost of distillation remains notoriously difficult, as it has decreased dramatically several times in the past decade~\cite{litinski2019magic, gidney2019efficient}. Even during the preparation of this manuscript, two new protocols were proposed that again promise substantial reductions in distillation cost~\cite{wills2024constant, gidney2024magic}.

To allow future refinements, we report all resource estimates in terms of space-time volume (the product of execution time and the total number of qubits), which serves as a flexible proxy for total computational effort. This quantity can be easily updated as improved distillation or gate protocols reduce overhead.

\subsection{Scalability Model}
\label{ssec:scalability_model}

Although sub-threshold quantum computation has long been demonstrated~\cite{fowler2012surface}, only one experiment as of this writing has achieved error suppression beyond $d=3$~\cite{google2023suppressing}. Even then, the results did not exhibit the exponential suppression of logical error rate predicted by ~\cref{eq:exponential_suppression}. Such deviations from ideal scaling characterize the early fault-tolerant quantum computing regime~\cite{fellous2021limitations, katabarwa2023}.

To capture these limited quantum error correction capabilities, Katabarwa et al.~\cite{katabarwa2023} proposed several error models 
in which the physical error rate increases with the size of the system.
The first model is the~\emph{power law} model:
\begin{align}
\label{eq:scalability_power_law}
    p_{\text{phys}} & = p_0 n^{\frac{1}{s}}
\end{align} 
and the second is the~\emph{logarithmic} model:
\begin{align}
\label{eq:scalability_logarithmic}
    p_{\text{phys}} &= p_0\left( 1 + \frac{1}{s}\ln{n}\right)
\end{align} 
where $p_0$ is the single qubit error rate for the corresponding hardware architecture, $n$ is the number of physical qubits, and $s$ denotes the~\emph{scalability} of the device. 
The power model offers a more conservative view of scalability, while the logarithmic model, being more optimistic, suggests a slower increase in error rates with system size. In Section~\ref{sec:min_scalability}, we explore the implications of both models for resource estimation and show that the effect of scalability is largely independent of the specific model used.

The concept of scalability arises from the presence of scale-dependent errors, driven by engineering challenges that intensify as system size grows. For a given finite scalability parameter ($s < \infty$), there exists an upper bound on the problem size that a given architecture can support~\cite{katabarwa2023}. The models in \cref{eq:scalability_power_law}-\ref{eq:scalability_logarithmic} quantify this relationship. For high-speed \emph{type B} devices, scalability constraints may reflect frequency-crowding errors~\cite{Hutchings2017}, while for high-fidelity \emph{type A} devices, they may capture cross-mode coupling effects~\cite{Leung2018, Liang2024}.

Rather than specifying exact microscopic mechanisms for each model, we adopt the broader view that scalability---regardless of its physical origin---will continue to constrain performance in the near term. Katabarwa et al. applied these models to a simplified computational setting, demonstrating that scalability constraints can cap achievable problem sizes. Here, we extend that analysis to realistic quantum chemistry applications, evaluating how finite scalability limits the accessible problem size for different hardware classes and how these constraints inform hardware design for industrially relevant catalytic systems.

\subsection{Open-shell Catalysts Instances}
\label{ssec:instances}

To conduct a comprehensive scalability analysis, we selected a diverse set of open-shell systems from Neugebauer et al.~\cite{neugebauer2023}, focusing on instances that capture a range of chemical complexity and relevance to electrocatalytic applications. The selected instances include:

\begin{itemize}
    \item \textbf{Instances 3, 4, and 6:} Transition-metal metallocenes that serve as model systems for redox processes, owing to their well-defined oxidation states and reversible redox behavior. These compounds typically feature cyclopentadienyl ligands and exhibit distinct electronic properties governed by their metal centers.

    \item \textbf{Instances 13 and 14:} Additional metallocenes with varying spin multiplicities and electronic configurations, enabling analysis of how metal–ligand interactions influence catalytic activity and stability.

    \item \textbf{Instances 22 and 23:} Cobalt-based complexes that play key roles in proton reduction and hydrogen evolution reactions, representing systems with strong electron correlation and multi-electron redox processes.

    \item \textbf{Instances 27 and 28:} Planar complexes containing redox-active ligands, which pose additional challenges due to possible non-innocent ligand behavior and are important prototypes for hydrogen-generation catalysis.

\end{itemize}

For each molecule, we considered two charge states---neutral and oxidized---relevant to calculating vertical ionization potentials. Such calculations are particularly demanding for classical electronic-structure methods because the oxidized species are not at their equilibrium geometries, which may lead to increased multi-reference character and breakdown of commonly used single-reference approximations. Accordingly, our instance set was designed to span a range of electronic structures of varying levels of complexity.

By applying the same active-space selection protocol used by Neugebauer et al.~\cite{neugebauer2023}, we ensured that the static correlation effects in the ground states are captured accurately and that the problem sizes are representative of the realistic quantum chemistry computations for the first-row transition metal systems.

\begin{table}[!h]
    \centering
\begin{tabular}{rlrrllrrr}
\hline\hline
 Instance &  Type &  $Z_{ox}$ &  $Z_{in}$ & $S_{in}$ & $S_{ox}$ &  $n_{e,ox}$ &  $n_{e,in}$ &  $n_o$ \\
\hline
        3 &    SR &          1 &          0 &    1 &  3/2 &     31 &     32 &  34 \\
        4 &    SR &          1 &          0 &    0 &  1/2 &     33 &     34 &  34 \\
        6 &    SR &          2 &          1 &    0 &  1/2 &     35 &     36 &  34 \\
       13 & SR/MR &          1 &          0 &  1/2 &    0 &     34 &     35 &  34 \\
       14 & SR/MR &          1 &          0 &    1 &  1/2 &     29 &     30 &  43 \\
       22 & SR/MR &          1 &          0 &    0 &  1/2 &     31 &     32 &  32 \\
       23 &    MR &          1 &          0 &    0 &  1/2 &     11 &     12 &  10 \\
       23 &    MR &          1 &          0 &    0 &  1/2 &     27 &     28 &  36 \\
       23 &    MR &          1 &          0 &    0 &  1/2 &     63 &     64 &  64 \\
       27 &    MR &          0 &         -1 &    1 &  1/2 &     15 &     16 &  16 \\
       27 &    MR &          0 &         -1 &    1 &  1/2 &     43 &     44 &  45 \\
       27 &    MR &          0 &         -1 &    1 &  1/2 &     61 &     62 &  74 \\
       28 &    MR &          0 &         -1 &  1/2 &    0 &      8 &      9 &  14 \\
       28 &    MR &          0 &         -1 &  1/2 &    0 &     44 &     45 &  45 \\
       28 &    MR &          0 &         -1 &  1/2 &    0 &     62 &     63 &  74 \\
\hline
\end{tabular}
    \caption{Parameters of the problem instances and their multi-reference character according to Neugebauer et al.\cite{neugebauer2023}. $Z$: molecule charge, $S$: spin quantum number, $n_e$: number of active electrons, $n_o$: number of active orbitals. We used subscripts `in' and `ox' to denote initial and oxidized states, respectively. For the particularly challenging subclass of multi-reference (MR) problems, we considered three active spaces per instance.}
    \label{tab:instance_par}
\end{table}

The chemical diversity and electronic complexity of these systems provide a robust basis for evaluating scalability across different computational regimes and serve as representative case studies throughout this analysis.

In this work, we employ double-factorized QPE~\cite{Q_catalysis_resources, lee_even_2021} to estimate quantum resources required for computing vertical ionization potentials of the model electrocatalysts, both with and without scalability constraints. QPE is a fundamental quantum algorithm for determining eigenvalues of a unitary operator and underpins many quantum-chemistry applications, including ground-state energy estimation and excited-state spectroscopy. The ``double factorized'' variant utilizes a block-encoding technique that exploits the sparsity of two-body interactions in the molecular Hamiltonian, thereby reducing the computational cost of QPE while maintaining chemical accuracy.

\section{Minimum Scalability for Catalysis Instances}
\label{sec:min_scalability}

Scalability characterizes the ability of quantum hardware to maintain low error rates as its size increases. In this study, we use the infinite-scalability model to represent the ideal fault-tolerant limit and the finite-scalability model to describe early fault-tolerant devices where error suppression is constrained by engineering limits. Within the QPE framework, the primary goal is to determine the optimal number of physical qubits and circuit depth required to achieve a sufficiently low logical and physical error rate for algorithmic success. Having a sufficiently high scalability is therefore crucial for executing a quantum algorithm of a given problem size.

\begin{figure}[!ht]
\center
\includegraphics[width=1.0\linewidth]{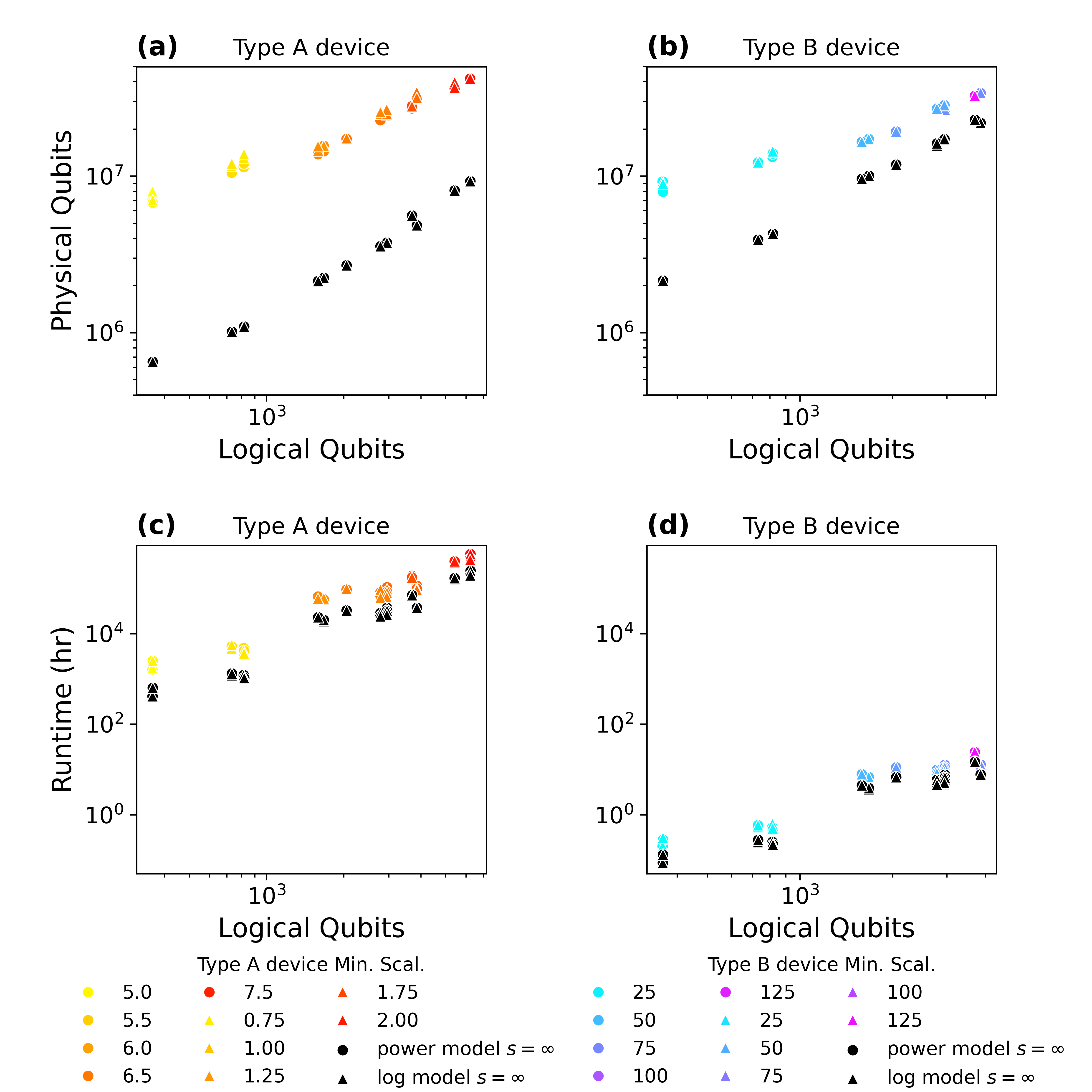}
\caption{The effect of introducing scalability on physical qubit and runtime requirements for catalysis instances. Round dots represent the power law model defined in \cref{eq:scalability_power_law}, while triangular dots correspond to the logarithmic model defined in \cref{eq:scalability_logarithmic}.}
\label{fig:logical_comparison}
\end{figure}

We evaluate the minimum scalability $s_\text{min}$ required for open-shell catalysis instances presented in \cref{ssec:instances} on both high-fidelity, slower devices (\emph{type A}) and high-speed, lower-fidelity devices (\emph{type B}). Our analysis examines how incorporating finite-scalability models impacts QPE resource estimates. Following Katabarwa et al., the problem size is measured in terms of the number of logical qubits. Figure \ref{fig:logical_comparison} shows how finite scalability modifies physical resource requirements across systems of different sizes.

\begin{figure}[!ht]
\center
\includegraphics[width=0.8\linewidth]{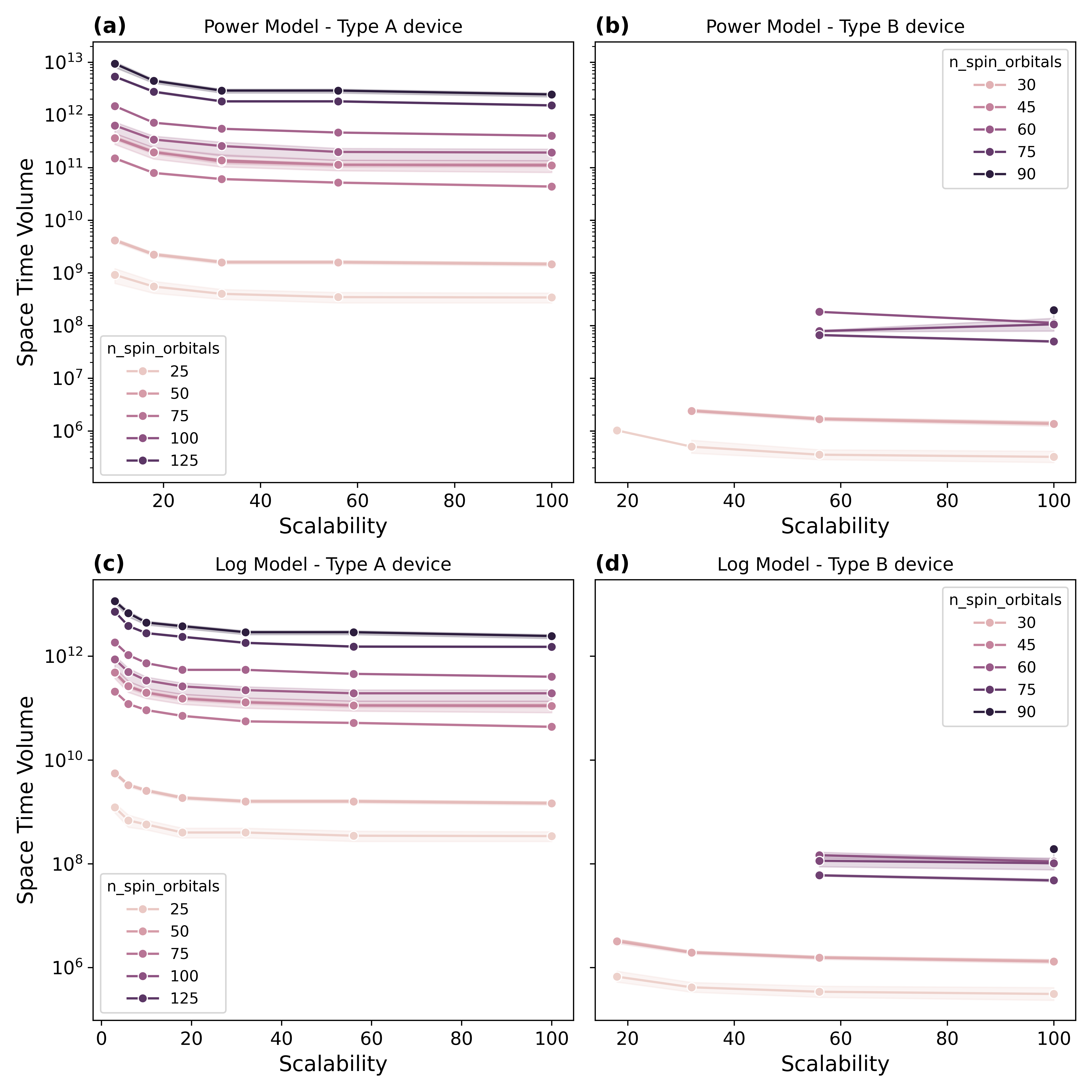}
\caption{The effect of introducing scalability on the space-time volume for catalysis instances. Different colors represent systems of varying sizes. Missing data points in subplots (b) and (d) indicate cases where \emph{type B} devices failed, as the scalability value corresponding to that system size was below the minimum scalability requirement.}
\label{fig:scalability_effect}
\end{figure}

The first two subplots in \cref{fig:logical_comparison} show logical versus physical qubit counts, while the remaining two depict logical qubits versus runtime. Black markers denote the infinite-scalability limit, while colored markers represent finite-scalability data. Across all systems, incorporating finite scalability produces a systematic shift toward higher resource requirements, while the overall scaling trends remain consistent. This shift is more evident in smaller systems and gradually less pronounced for larger ones, consistent with the mathematical behavior of the scalability models. Importantly, the resulting increase in resources remains bounded: finite scalability raises qubit counts and runtimes but not by orders of magnitude large enough to invert the relative difficulty between small and large systems. This trend is illustrated in \cref{fig:scalability_effect}, where, for each system size, the rate of change in total space-time volume (runtime $\times$ number of physical qubits) decreases with increasing scalability and saturates within roughly one order of magnitude.

Quantitatively, \cref{fig:logical_comparison} shows that incorporating scalability in the \emph{type A} model leads to an average 7-fold increase in the number of physical qubits and a 3-fold increase in runtime. For the \emph{type B} architecture, the corresponding increases are roughly 2-fold and 1.8-fold, respectively. These results highlight the significant effect of finite scalability on both the physical resource requirements and operational efficiency of quantum computing devices.

To test model dependence, we repeated the minimum scalability $s_{min}$ requirements analysis using both the power-law and logarithmic scalability models introduced in \cref{ssec:scalability_model} across the same set of instances. 
In Fig.~\ref{fig:logical_comparison}, we present data for \emph{type A} qubits on the left column and \emph{type B} qubits on the right. Round dots correspond to the power-law model [Eq.~\ref{eq:scalability_power_law}], while triangles represent the logarithmic model [Eq.~\ref{eq:scalability_logarithmic}].
The results confirm that, although derived from different mathematical forms, the two scalability models produce nearly identical outcomes for runtime and physical qubit requirements. Differences are most pronounced at smaller minimum scalability values, where \emph{type A} systems exhibit slightly higher variation in required resources between models. For \emph{type B} systems, which operate at larger minimum scalabilities, the distinction between models becomes negligible. This convergence arises naturally from the asymptotic behavior of both models, as illustrated in \cref{fig:scalability_effect}, where the space-time volume plateaus at large $s$. The near-overlap of data from the two models confirms that the conclusions are insensitive to the specific mathematical form chosen for scalability, underscoring the robustness of our analysis.

Next, we determine the minimum hardware scalability necessary for accurate quantum-chemistry simulations. For reference, IBM’s hardware (as of the time of this draft, 2023–2024) exhibits a scalability of approximately 1.75 with an error rate of $p_0 = 0.005$~\cite{katabarwa2023}. As the chemical system size increases, the corresponding number of logical qubits rises, leading to higher minimum scalability requirements, as shown in \cref{fig:logical_comparison}. For the smallest catalytic instance under the power-law model, the required scalability is roughly 5.0 for high-fidelity \emph{type A} devices and below 25 for high-speed \emph{type B} devices. For the largest systems tested, the corresponding values reach 7.5 for \emph{type A} and 125 for \emph{type B} architectures, with comparable trends for the logarithmic model. These results indicate that high-fidelity \emph{type A} hardware requires substantially lower scalability to compute the same problem size as high-speed \emph{type B} hardware, primarily due to the lower error rates achievable in \emph{type A} devices~\cite{Brown2011, Li2023}.

Overall, our findings show that incorporating finite scalability into resource estimation significantly refines the projected hardware requirements without qualitatively altering scaling trends. While resource demands increase when scalability limits are included, these increases remain bounded, and the analysis is largely independent of the chosen model. Notably, high-fidelity, slower \emph{type A} hardware can accommodate larger catalytic systems at lower scalability thresholds. For a given \emph{type A} gate that is roughly three orders of magnitude slower than a corresponding high-speed \emph{type B} gate performing the same operation, it is natural to ask whether there exist scenarios in which \emph{type A} devices could still compete with \emph{type B} hardware in total runtime. We explore this question in detail in \cref{sec:competitive_analysis}.

Over time, hardware development will increasingly focus on enhancing scalability, which consequently enables the solution of progressively more complex problem instances.
Tracking hardware progress through scalability provides a quantitative way to identify which architectures will reach larger problem sizes first. Our results indicate that devices with lower logical error rates, and therefore lower scalability requirements for a given problem, will achieve this milestone sooner. This shows that infinite scalability, while idealized, is not a prerequisite for quantum advantage. Sub-optimal but improving scalability can already support meaningful large-scale simulations, suggesting that chemically relevant quantum computations may become feasible earlier than previously anticipated.

\section{Competitive Analysis} \label{sec:competitive_analysis}

Having established that high-fidelity, lower-speed \emph{type A} architectures can handle larger problem sizes at lower minimum scalability than high-speed, lower-fidelity \emph{type B} architectures, we now examine the practical implications of these findings.
Notably, a large time-scale disparity ($t$) exists between \emph{type A} and \emph{type B} gates performing identical operations, with certain \emph{type A} gates being roughly three orders of magnitude slower.

To compensate for this gate-time disparity, we assume that \emph{type A} architectures can exploit parallelization to accelerate computation.
This parallelization is modeled using the game of surface codes framework~\cite{litinski2019game}, in which computational time is measured in units of tocks ($t_L$), corresponding to groups of $d$ surface-code cycles:
\begin{align}
C &=  t_{L} d.
\end{align}
Thus, when one is compiling using the game of surface codes, one can think of the total volume of the computation $V$ to be held constant and the space-time trade-off between logical qubits $k$ and tocks $t_L$ can be written as:
\begin{align}
\label{eq:space-time_volume}
V &\approx k_B t_{\text{L,B}} \approx k_A t_{\text{L,A}} 
\end{align} 
The related variables and their specifications in the quantum circuit across architectures are defined in table \cref{tab:vars}. 

In general, as scalability increases, a higher code distance $d$ becomes feasible for each logical qubit, allowing it to better 'overcome' errors.
Therefore, $d$ is restricted to the range $1$–$100$, since larger code distances would lead to impractically long logical cycle times.

To rigorously assess the conditions under which high-fidelity \emph{type A} devices might competitively align with high-speed \emph{type B} devices, we aim to find a regime where the total time $T$ of both technologies has a ratio that approaches or is less than approximately $10$:
\begin{align}
\frac{T_A}{T_B} &= \frac{t_{A} \cdot C_A}{t_{B} \cdot C_B} =  \frac{t_A}{t_B} \cdot \frac{d_A}{d_B}  \cdot \frac{t_{\text{L,A}}}{t_{\text{L,B}}} 
\label{eq:first_pass}
\end{align}
where $t$ is the time scale for gates, $d$ is the code distance, $t_L$ is the number of tocks in the computation~\cite{litinski2019game}, and $C$ is the number of surface code cycles per tock. 

\begin{table}[tb]
    \centering
    \caption{Reference Table of Variables and Their Specifications in the Quantum Circuit Across Architectures}
    \label{tab:vars}
    \begin{tabular}{c c l}
    \hline\hline
    \textbf{Variable} & \textbf{Value} & \textbf{Description} \\
    \hline
    $C$ & & Number of logical cycles \\
    $n$ & & Number of physical qubits\\
    $k$ & & Number of logical qubits \\
    $d$ & & Code distance \\
    $T$ & & Total time \\
    $s$ & & Scalability \\
    \hline
    $t_{A}$ & $10^{-4}$ & Gate time for \emph{type A} device, unit: seconds.\\
    $t_{B}$ & $10^{-7}$ & Gate time for \emph{type B} device, unit: seconds.\\
    \hline
    $p_{0;\text{A}}$ & $10^{-4}$ & Single qubit error rate for \emph{type A} device architecture\\
    $p_{0;\text{B}}$ & $10^{-3}$ & Single qubit error rate for \emph{type B} device architecture\\
    \hline
    $t_{B} / t_{A}$ & $10^{-3}$ & Gate time ratio \\
    \hline
    $d_{A}$, $d_{B}$ & $\leq 100$ & $d_{A} = d$; $d_{B} = d + d'$; $d_{B} \geq d_{A}$\\
    \hline
    $K_{B}$ & $\approx 10^3$ & Equivalent to the number of qubits in the circuit \\
    $K_{A}$ & $\geq K_{B}$ & \\
    \hline
    $p_{th}$ & $0.01 - 0.1$ & Error threshold; typically targeted between 1\% and 10\%\\
    \hline 
    \end{tabular}
\end{table}

Substituting \cref{eq:space-time_volume} into \cref{eq:first_pass} expresses the total-time ratio in terms of the circuit space-time volume $V$ and the number of logical qubits $k$:
\begin{align}
    \frac{T_A}{T_B} = \frac{t_A}{t_B} \cdot \frac{d_A}{d_B}  \cdot \frac{V/k_{A}}{V/k_{B}} 
\end{align}

For surface-code implementations, the space-time volume of high-speed \emph{type B} devices can be expressed by combining the power-law scalability model [Eq.~\ref{eq:scalability_power_law}] with the standard logical-error formulation~\cite{katabarwa2023,fowler2019}:
\begin{align}
\label{eq:space_time_volume_and_phys_error_rate}
    \frac{1}{V} = p_{L} = O \cdot (\frac{p_{\text{phys}}}{p_{\text{th}}})^{\frac{d+1}{2}}.
\end{align}

In \cref{eq:space_time_volume_and_phys_error_rate}, $p_L$ denotes the logical error rate, which corresponds to the inverse of the space-time volume $1/V$, implying that each unit of $V$ carries an associated probability of failure. $O$ denotes the error overhead required for fault-tolerant computation, and $O = 0.1$ is used in this paper ~\cite{goings2022, fowler2019}. 
$p_{\text{phys}}$ is the physical error rate and $p_{\text{th}}$ is the error threshold of the fault-tolerant protocol. By incorporating the power law model for $p_{\text{phys}}$, we derive the following expression:
\begin{align}
\label{eq:logical_e_rate_2}
    \frac{1}{V} &= O \cdot (\frac{p_0 \cdot n^{\frac{1}{s}}}{p_{\text{th}}})^{\frac{d+1}{2}}.
\end{align} 
In this formula, the number of physical qubits, $n$, can be expressed as a function of $d$, which comes from the size of a single logical qubit multiplied by the number of logical qubits, $k$ \cite{katabarwa2023, goings2022}:
\begin{align}
    n = 2 \cdot (d+1)^2 \cdot k.
\end{align}
Consequently, $n$ in ~\cref{eq:logical_e_rate_2} can be replaced with $d$ and $k$, leading us to the following expression for $1/V$: 
\begin{align}
\label{eq:logical_e_rate_4}
    \frac{1}{V} = O \cdot \left[ (\frac{p_0}{p_{\text{th}}} \cdot (2k)^{\frac{1}{s}})^{\frac{1}{2}} \cdot (d+1)^{\frac{1}{s}}\right] ^{d+1}.
\end{align} 

Therefore, the total time $T$ for high-speed \emph{type B} devices utilizing the surface code can be obtained using the equation below:
\begin{align}
\begin{split}
\label{eq:T_B_expression_for_LDPC}
    T_B &= t_B \cdot d_B \cdot \frac{V}{k_B}\\
    &= \frac{t_B \cdot d_B}{k_B} \cdot O^{-1} \cdot \left[ (\frac{p_0}{p_{th}} \cdot (2 k_B)^{\frac{1}{s_B}})^{\frac{1}{2}} \cdot (d_B+1)^{\frac{1}{s}}\right] ^{-(d_B+1)}.
\end{split}
\end{align}

\subsection{Surface Code Comparison}

For the runtime comparison, we first derive a closed-form expression for the code distance $d$ and then substitute it into the scalability model. To solve for $d$ as a function of scalability $s$, we raise \cref{eq:logical_e_rate_2} to the power of $s$ on both sides of the equation.
\begin{align}
    1 = {(V*O)}^s \cdot \left[ \left[(\frac{p_0}{p_{\text{th}}})^s \cdot (2k)\right]^{\frac{1}{2}} \cdot (d+1) \right] ^{d+1}.
\end{align}
The above expression can be further simplified by defining variables:
\begin{align*}
\label{eq:define_a}
a &:= \left[2k \left(\frac{p_0}{p_{\text{th}}}\right)^s\right]^{\frac{1}{2}},\\
b &:= (VO)^{-s},\\
d' &:= (d + 1).
\end{align*}
Using these definitions, we derive the following equation: 
\begin{align}
    b = \left(a{d'}\right)^{d'}.
\end{align}

By utilizing the properties of the natural logarithm, we can transform the above equation into the form: $a \cdot \ln(b) = e^{\ln(ad)} \cdot \ln(ad)$, making $\ln(ad)$ solvable in terms of the Lambert W function \cite[Equation 1.1]{mezo2022_lambertW}:
\begin{align}
\label{eq:lambert_w}
    \ln(a \cdot {d'}) = W\left(a \cdot \ln(b)\right)
\end{align}
This transformation allows for an explicit expression for $d'$, which can be further simplified based on the property of $W$ \cite[Equation 1.20]{mezo2022_lambertW}: 
\begin{align}
\label{eq:d'}
    d' &= \frac{1}{a} e^{W (a \ln{b})} \\
       &= \frac{\ln{b}}{W (a \ln{b})}.
\end{align}

The ratio between the total computational time in ~\cref{eq:first_pass} thus can be written in terms of $a$, $b$ and $d'$ as following:
\begin{align}
    \frac{T_{A}}{T_{B}} &= \frac{t_{A}}{t_{B}}\cdot \frac{\left({a_{A}}^{-1} \cdot e^{W\left(a_{A} \cdot \left(\ln{b_{A}}\right)\right)} - 1\right)}{\left({a_{B}}^{-1} \cdot e^{W\left(a_{B} \cdot \ln{\left(b_{B}\right)}\right)} - 1\right)} \cdot \frac{k_{B}}{k_{A}} \\
    &= \frac{t_{A}}{t_{B}}\cdot \frac{\left({\ln{(b_{A})}} / {W (a_{A} \cdot \ln{(b_{A})})} - 1\right)}{\left({\ln{(b_{B})}} / {W (a_{B} \cdot \ln{(b_{B})})} - 1\right)} \cdot \frac{k_{B}}{k_{A}}
\label{eq:T_ratio}
\end{align}

\subsection{LDPC Codes Comparison}
To give high-fidelity \emph{type A} qubits the best possible chance of competing with high-speed \emph{type B} qubits, we consider a model in which \emph{type A} employs a BB LDPC code (\cref{subsec:LDPC}).
We can describe the spacetime volume that can be executed on an LDPC code using the fit formula \cref{eqn:LDPC_logical_error_rate} from ~\cite{bravyi2024high}, and calculate the corresponding total time $T$ for this type of device:

\begin{align}
\begin{split}
\label{eq:T_A_expression_for_LDPC}
    T_A &= t_A \cdot d_{\text{circ}} \cdot \frac{V}{k_A}\\
    &= \frac{t_A \cdot d_{\text{circ}}}{k_A} \cdot {p_{\text{phys,A}}}^{-\frac{d_{\text{circ}}}{2}} \cdot \exp\left(-c_0 - c_1 \cdot p_{\text{phys,A}} - c_2 \cdot {p_{\text{phys,A}}}^2 \right) \\
    &= \frac{t_A \cdot d_{\text{circ}}}{k_A} \cdot \left(p_{0,\mathrm{A}} \cdot n^{\frac{1}{s_A}}\right)^{-\frac{d_{\text{circ}}}{2}} \cdot \exp\left(-c_0 - c_1 \left(p_{0,\mathrm{A}} \cdot n^{\frac{1}{s_A}}\right) - c_2 \left(p_{0,\mathrm{A}} \cdot n^{\frac{1}{s_A}}\right)^2 \right)
\end{split}
\end{align}
where $d_{\text{circ}}$ denotes the number of error-correction rounds performed between successive operations. For lattice surgery, one would expect that $d_{\text{circ}} = d$, but for transversal gates it should only be a constant. $c_0$, $c_1$, and $c_2$ are constants obtained via fitting which was done in table 3 of the supplementary material in~\cite{bravyi2024high}. A recent paper done by Quera says that $d_{\text{circ}} = 3$ for decoders that we expect to scale~\cite{cain2024correlated}.

\subsection{Competitiveness Results}

We study when high-fidelity \emph{type A} devices can achieve runtime competitiveness with high-speed \emph{type B} devices using the representative instance $14_{\mathrm{in}}$, which has the largest estimated space-time volume in our set (on the order of $10^{10}$). Our goal is to find parameter regimes where $1 \le T_A/T_B \le 10$.
We first fix physically reasonable parameter values for gate times, physical error rates, and thresholds for both architectures (Table~\ref{tab:vars}). We then scan scalability ranges for \emph{type A} and \emph{type B}, each spanning $1$ to $100$. For every pair $(s_A, s_B)$, we evaluate whether the total-time ratio falls within the target band using \cref{eq:T_ratio}, \eqref{eq:T_B_expression_for_LDPC}, and \eqref{eq:T_A_expression_for_LDPC}.

\textbf{Type A with surface code: } computing $T_A/T_B$ requires valid code distances $d_A$ and $d_B$ and a specification of the logical-qubit ratio $k_A/k_B$. We constrain code distances by $d<100$ for physical relevance and set the \emph{type B} logical-qubit budget to $k_B = 10^3$. For each $(s_A, s_B)$, we compute $d_A$ and $d_B$ from the inequality
\begin{align}
\label{eq:code_distance_search}
    \frac{1}{V} \geq \left[ (\frac{p_{0}}{p_{\text{th}}} \cdot (2k)^{\frac{1}{s}})^{\frac{1}{2}} \cdot (d+1)^{\frac{1}{s}}\right] ^{d+1},
\end{align} 
using the closed-form Lambert-$W$ solution, and then solve numerically for the smallest $k_A$ that yields a total runtime ratio within the target band. When a feasible solution exists, we record $k_A/k_B$ as the degree of runtime competitiveness.

\begin{figure}[ht]
\center
\includegraphics[width=0.7\linewidth]{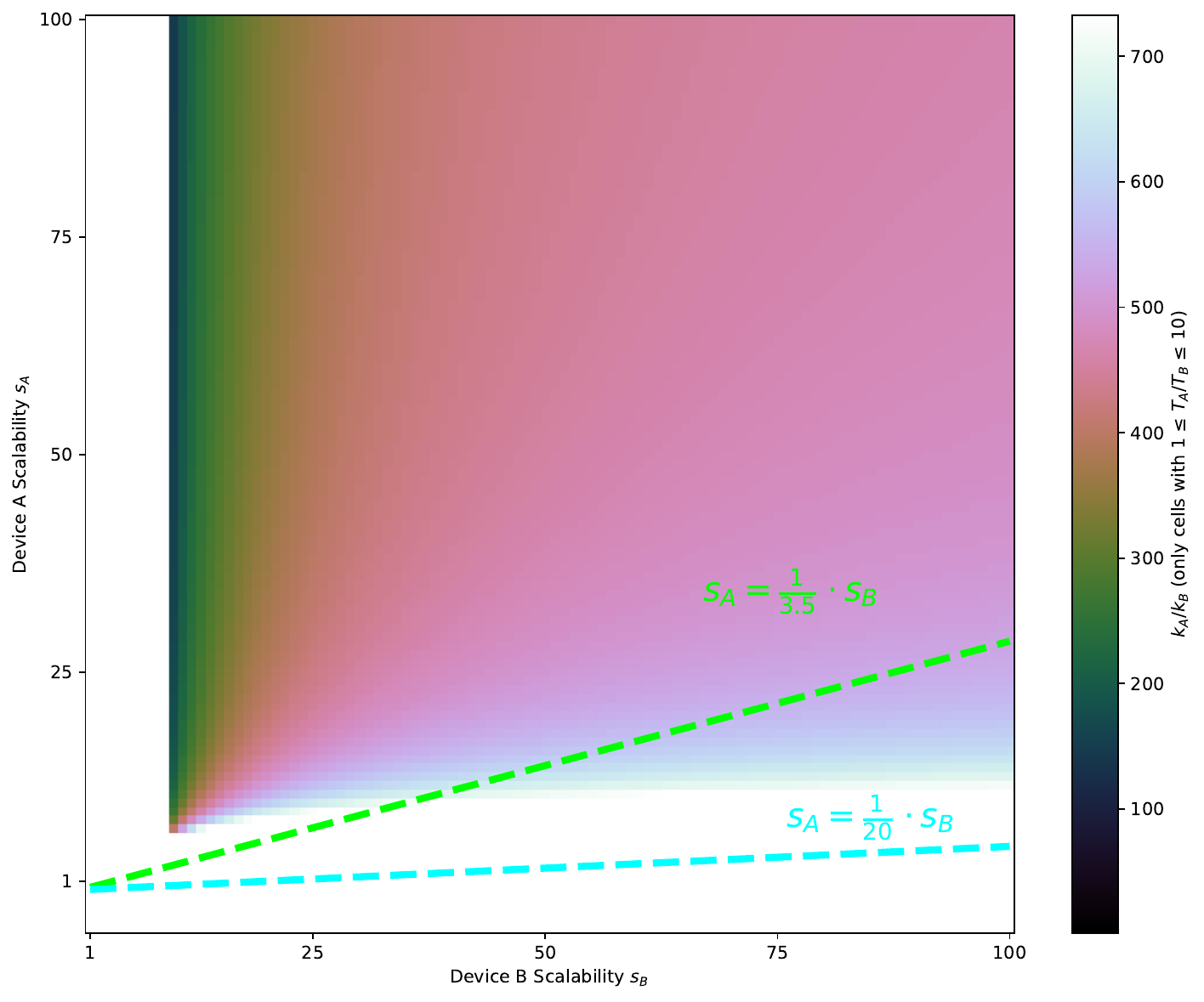}
\caption{The ratio of qubits required for \emph{type A} hardware to achieve competitive runtime with \emph{type B} hardware. The x-axis represents the scalability of \emph{type B} qubits, while the y-axis represents the scalability of \emph{type A} qubits. 
The color gradient in the plot represents the ratio of \emph{type A} logical qubits ($k_{A}$) to \emph{type B} logical qubits ($k_{B}$), indicating the level where competitiveness is achieved. The green and blue lines indicate the highest and lowest scalability ratios found for the catalysis instances.}
\label{fig:competitiveness_heatmap}
\end{figure}

We repeat this search over the full $(s_A, s_B)$ grid. The resulting competitiveness regions appear in \cref{fig:competitiveness_heatmap}, where the colored area marks parameter pairs for which \emph{type A} achieves runtime competitiveness; the color scale encodes $k_A/k_B$. The horizontal white band at the bottom of \cref{fig:competitiveness_heatmap} marks regions where the \emph{type A} technology fails to achieve runtime competitiveness. The vertical white strip on the left arises where $s_B$ is too small for any $d_B \le 100$ to satisfy \eqref{eq:code_distance_search}.

\begin{figure}[ht]
\center
\includegraphics[width=0.7\linewidth]{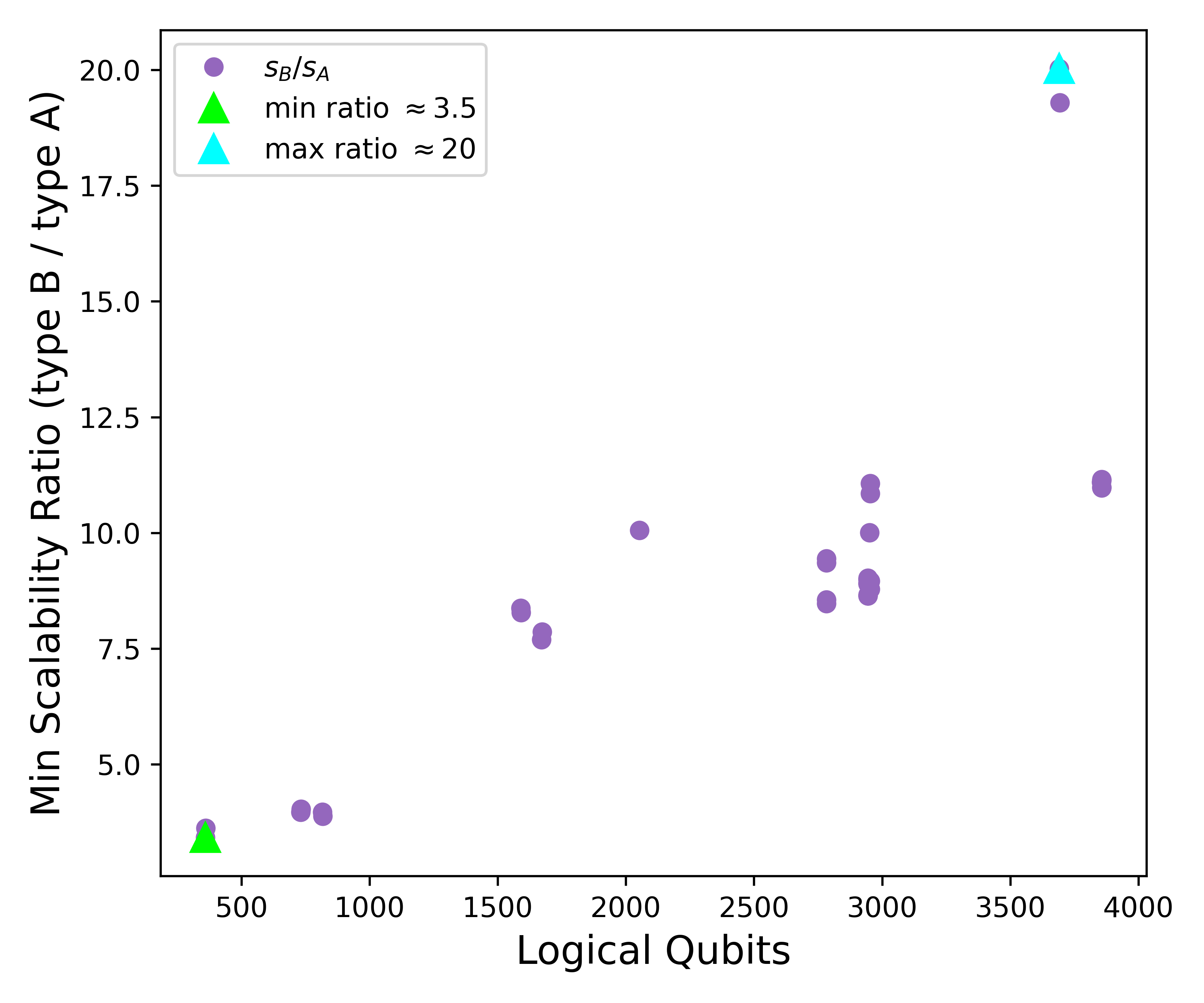}
\caption{The minimum scalability ratio (purple dots) between \emph{type B} ($s_{B}$) and \emph{type A} ($s_{A}$) hardware architectures for the catalysis instances, highlighted with green and blue triangles for the lowest and highest ratios, respectively.}
\label{fig:min_scalability_ratio}
\end{figure}

Section \ref{sec:min_scalability} showed that \emph{type A} requires substantially lower minimum scalability than \emph{type B} for identical instances. We summarize the minimum-scalability ratio $s_{\min;B}/s_{\min;A}$ as a function of problem size in \cref{fig:min_scalability_ratio}; it ranges from about 3.5 to 20 and generally increases with system size.

We overlay these extremes as dashed lines with corresponding colors in \cref{fig:competitiveness_heatmap}. 
The plot demonstrates that smaller $s_{min;B}/s_{min;A}$ ratios, corresponding to smaller problem sizes, allow for a competitive regime to be achieved for high-fidelity \emph{type A} devices with a reasonable number of qubits. 
However, as $s_{min;B}/s_{min;A}$ ratio increases with larger problem sizes, the competitive regime shifts into the area marked by the blue dashed line, indicating that achieving runtime competitiveness is not feasible with physically meaningful parameters for larger problem sizes.

Overall, \cref{fig:competitiveness_heatmap} and \cref{fig:min_scalability_ratio} indicate that while high-fidelity \emph{type A} devices can initially handle larger problems more efficiently in terms of minimum scalability, they lose runtime competitiveness as problem size increases and as scalability requirements for \emph{type B} improve. 
Building on this, we next assess whether advanced code families can further extend the competitiveness of \emph{type A} architectures beyond the surface-code limit.

\begin{figure}[ht]
\center
\includegraphics[width=0.7\linewidth]{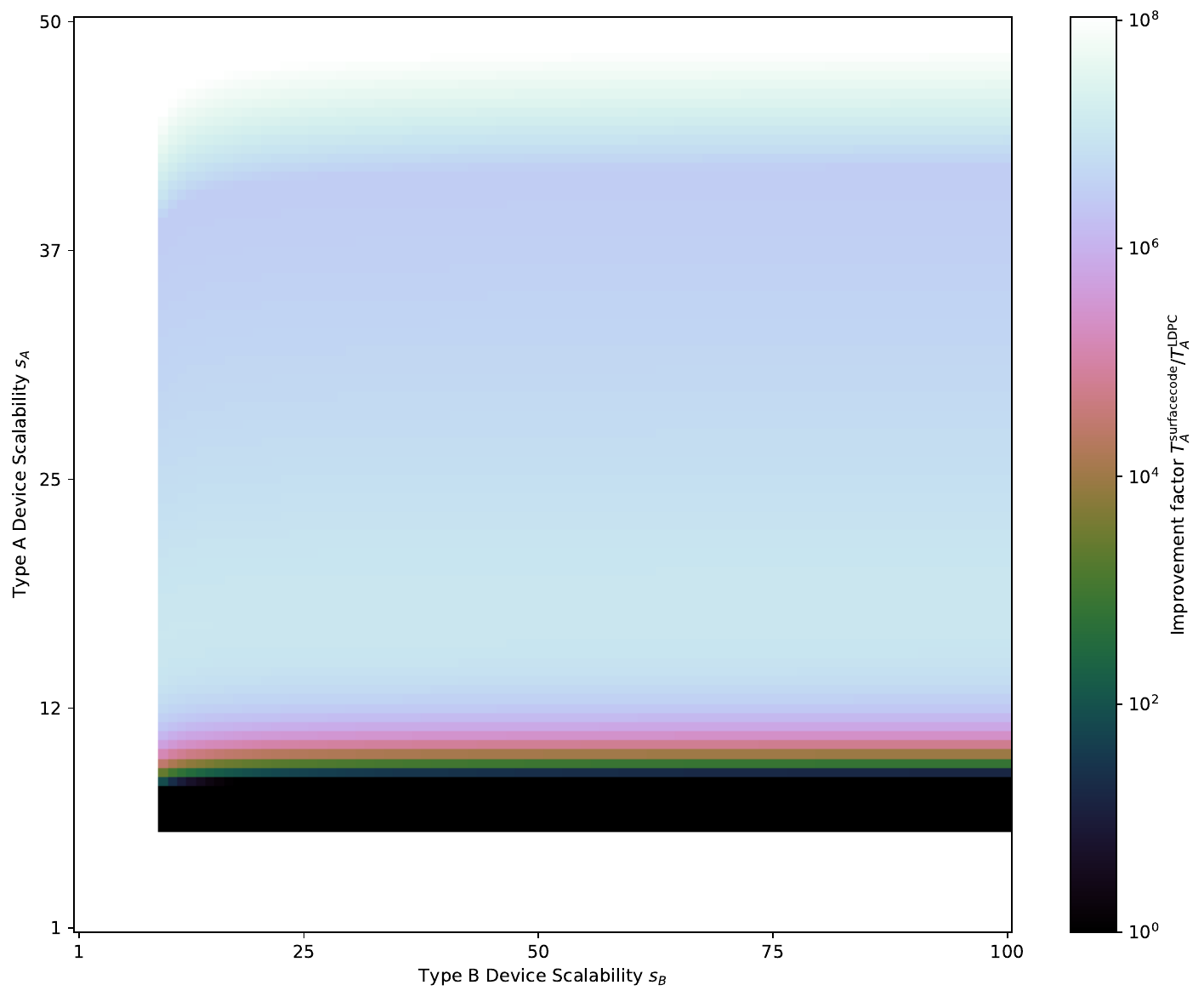}
\caption{Heatmap of the improvement factor $T_A^{\mathrm{surface code}}/T_A^{\mathrm{LDPC}}$ over the $(s_A,s_B)$ grid. For each point we first pick $k_A$ using the surface-code model under the band policy $1 \le T_A/T_B \le 10$, then hold that same $k_A^\star$ and evaluate the best LDPC option.}
\label{fig:competitiveness_heatmap_LDPC}
\end{figure}

\textbf{Type A device with LDPC code}: to visualize how LDPC-based \emph{type A} architectures perform relative to surface codes, we compute the ratio $T_A^{\mathrm{LDPC}} / T_A^{\mathrm{SC}}$ across a grid of scalability parameters $(s_A, s_B)$.
The resulting heatmap \cref{fig:competitiveness_heatmap_LDPC} quantifies how efficiently \emph{type A} device can execute logical operations when protected by modern LDPC codes compared to a surface-code-protected device. Values greater than unity indicate speed advantages for the LDPC architecture.

The improvement arises because LDPC codes relax the asymptotic space-time overhead that dominates surface-code performance. In surface codes, logical error suppression scales as $\exp(-\alpha d)$ but at the cost of quadratic physical qubit overhead in distance $d$, while LDPC codes achieve similar suppression with constant-weight parity checks and near-linear scaling in qubit count. As a result, for large $s_A$, representing lower-noise hardware, the effective logical clock cycle decreases much faster than in surface-code-limited architectures. The extreme high-ratio regime represents a limit where the LDPC code operates near its threshold under ideal decoding; realistic hardware or finite-rate effects would likely compress this to a much more modest improvement. Still, these results confirm that LDPC-protected \emph{type A} systems can, in principle, deliver lower logical runtime than surface-code-protected \emph{type A} devices.

Collectively, the results presented in this section clarify how scalability, code performance, and architectural characteristics jointly determine the practical competitiveness of selective platforms. Within the surface-code framework, high-fidelity \emph{type A} devices can temporarily offset slower gate speeds through parallelization and lower logical-error rates, but this advantage diminishes as problem size and scalability demands grow. Incorporating LDPC codes partially restores competitiveness by relaxing the quadratic space-time overhead intrinsic to surface codes, highlighting the importance of advanced codes in extending the useful lifetime of early high-fidelity architectures. Overall, these analyses underscore that sustained progress toward utility-scale quantum computation will rely on co-optimizing hardware fidelity, scalability, and code efficiency rather than pursuing any single improvement in isolation.

\section{Conclusion}
This work examined scalability as a defining constraint for EFTQC and its implications for quantum simulations of open-shell catalytic systems. 
Our investigation into the scalability of quantum systems underscores its impact on the feasibility and efficiency of executing complex quantum chemistry problems as we transition from NISQ to fully fault-tolerant regimes. The key findings can be summarized as follows:
\begin{itemize}
  \item Finite scalability model provides essential insights into the performance of early fault-tolerant architectures, which is vital for practical quantum computing applications.
  \item Incorporating scalability into quantum phase estimation significantly alters resource estimations, increasing the requirements for physical qubits and extending computation times.
    \item High-fidelity, lower-speed (\emph{type A}) architectures require significantly lower minimum scalability (below 8 for the largest catalytic systems studied) compared to high-speed, lower-fidelity (\emph{type B}) devices, which demand values above 100.
    \item While high-fidelity \emph{type A} architectures can achieve competitive runtimes with high-speed \emph{type B} systems at smaller problem sizes, this advantage diminishes as the number of logical qubits increases.
    \item Using LDPC codes instead of surface codes partially restores this competitiveness by reducing the quadratic space-time overhead, extending the feasible regime for high-fidelity architectures.
\end{itemize}

Together, these results demonstrate that infinite scalability, while idealized, is not required for utility-scale quantum computation. Systems with finite but improving scalability can achieve meaningful chemical simulations sooner than previously anticipated, especially when combined with optimized code architectures. 
This is particularly relevant in the EFTQC regime and underscores the value of explicitly modeling scalability to inform the strategic development of quantum computing technologies.

Future work should extend this framework to additional hardware platforms and error-correction schemes to broaden the understanding of how finite-scalability effects propagate across quantum technologies.
Such studies will help clarify how scalability constraints shape algorithmic feasibility and resource trade-offs in different architectures, guiding realistic assessments of early fault-tolerant quantum systems for realistic applications.

\section*{Acknowledgements}
The authors thank Nicole Bellonzi for her contributions and collaboration during the development of this work. 

\section*{Data availability \label{DAS}} 
The raw data supporting the conclusions of this article will be made available by the
authors on request.

\section*{Conflicts of interest \label{COI}} 
Yanbing Zhou and Athena Caesura were previously employed by Zapata Computing, Inc., and Xavier Jackson, Corneliu Buda, Clena M. Abuan, and Shangjie Guo were employed by bp plc during the course of this work.

\printbibliography

@article{fowler2012surface,
  title={Surface codes: Towards practical large-scale quantum computation},
  author={Fowler, Austin G and Mariantoni, Matteo and Martinis, John M and Cleland, Andrew N},
  journal={Physical Review A},
  volume={86},
  number={3},
  pages={032324},
  year={2012},
  publisher={APS}
}

@article{horsman2012surface,
  title={Surface code quantum computing by lattice surgery},
  author={Horsman, Dominic and Fowler, Austin G and Devitt, Simon and Van Meter, Rodney},
  journal={New Journal of Physics},
  volume={14},
  number={12},
  pages={123011},
  year={2012},
  publisher={IOP Publishing}
}

@article{de2020zx,
  title={The ZX calculus is a language for surface code lattice surgery},
  author={de Beaudrap, Niel and Horsman, Dominic},
  journal={Quantum},
  volume={4},
  pages={218},
  year={2020},
  publisher={Verein zur F{\"o}rderung des Open Access Publizierens in den Quantenwissenschaften}
}

@article{gidney2019efficient,
  title = {Efficient magic state factories with a catalyzed $\lvert CCZ\rangle \to 2 \lvert T\rangle$ transformation},
  author = {Gidney, Craig and Fowler, Austin G.},
  journal = {Quantum},
  volume = {3},
  pages  = {135},
  year   = {2019}
}

@article{litinski2019magic,
  title={Magic state distillation: Not as costly as you think},
  author={Litinski, Daniel},
  journal={Quantum},
  volume={3},
  pages={205},
  year={2019},
  publisher={Verein zur F{\"o}rderung des Open Access Publizierens in den Quantenwissenschaften}
}

@article{wills2024constant,
  title={Constant-Overhead Magic State Distillation},
  author={Wills, Adam and Hsieh, Min-Hsiu and Yamasaki, Hayata},
  journal={arXiv preprint arXiv:2408.07764},
  year={2024}
}

@article{cain2024correlated,
  title={Correlated decoding of logical algorithms with transversal gates},
  author={Cain, Madelyn and Zhao, Chen and Zhou, Hengyun and Meister, Nadine and Ataides, J and Jaffe, Arthur and Bluvstein, Dolev and Lukin, Mikhail D},
  journal={arXiv preprint arXiv:2403.03272},
  year={2024}
}

@article{gidney2024magic,
  title={Magic state cultivation: growing T states as cheap as CNOT gates},
  author={Gidney, Craig and Shutty, Noah and Jones, Cody},
  journal={arXiv preprint arXiv:2409.17595},
  year={2024}
}

@article{google2023suppressing,
  author  = {{Google Quantum AI}},
  title   = {Suppressing quantum errors by scaling a surface code logical qubit},
  journal = {Nature},
  volume  = {614},
  number  = {7949},
  pages   = {676--681},
  year    = {2023}
}

@article{katabarwa2023,
  title = {Early Fault-Tolerant Quantum Computing},
  author = {Katabarwa, Amara and Gratsea, Katerina and Caesura, Athena and Johnson, Peter D.},
  journal = {PRX Quantum},
  volume = {5},
  number = {2},
  pages = {020101},
  numpages = {20},
  year = {2024},
  month = {6},
  publisher = {American Physical Society},
  doi = {10.1103/PRXQuantum.5.020101},
  url = {https://link.aps.org/doi/10.1103/PRXQuantum.5.020101}
}

@article{litinski2019game,
  title={A game of surface codes: Large-scale quantum computing with lattice surgery},
  author={Litinski, Daniel},
  journal={Quantum},
  volume={3},
  pages={128},
  year={2019},
  publisher={Verein zur F{\"o}rderung des Open Access Publizierens in den Quantenwissenschaften}
}

@article{litinski2018lattice,
  title={Lattice surgery with a twist: simplifying clifford gates of surface codes},
  author={Litinski, Daniel and von Oppen, Felix},
  journal={Quantum},
  volume={2},
  pages={62},
  year={2018},
  publisher={Verein zur F{\"o}rderung des Open Access Publizierens in den Quantenwissenschaften}
}

@article{goings2022,
author = {Joshua J. Goings  and Alec White  and Joonho Lee  and Christofer S. Tautermann  and Matthias Degroote  and Craig Gidney  and Toru Shiozaki  and Ryan Babbush  and Nicholas C. Rubin },
title = {Reliably assessing the electronic structure of cytochrome P450 on today’s classical computers and tomorrow’s quantum computers},
journal = {Proceedings of the National Academy of Sciences},
volume = {119},
number = {38},
pages = {e2203533119},
year = {2022},
doi = {10.1073/pnas.2203533119},
URL = {https://www.pnas.org/doi/abs/10.1073/pnas.2203533119},
eprint = {https://www.pnas.org/doi/pdf/10.1073/pnas.2203533119}}

@misc{fowler2019,
      title={Low overhead quantum computation using lattice surgery}, 
      author={Austin G. Fowler and Craig Gidney},
      year={2019},
      eprint={1808.06709},
      archivePrefix={arXiv},
      primaryClass={quant-ph},
      url={https://arxiv.org/abs/1808.06709}, 
}

@article{neugebauer2023,
    author = {Neugebauer, Hagen and Vuong, Hung T. and Weber, John L. and Friesner, Richard A. and Shee, James and Hansen, Andreas},
    title = {Toward Benchmark-quality Ab Initio Predictions for 3d Transition Metal Electrocatalysts - A Comparison of CCSD (T) and ph-AFQMC},
    year = {2023},
    url = {https://chemrxiv.org/engage/chemrxiv/article-details/64dcee86dfabaf06ff5fa2d0},
    journal = {ChemRxiv}
}

@article{fellous2021limitations,
  title={Limitations in quantum computing from resource constraints},
  author={Fellous-Asiani, Marco and Chai, Jing Hao and Whitney, Robert S and Auff{\`e}ves, Alexia and Ng, Hui Khoon},
  journal={PRX Quantum},
  volume={2},
  number={4},
  pages={040335},
  year={2021},
  publisher={APS}
}

@misc{mezo2022_lambertW, 
    edition={1}, 
    title={The Lambert W Function: Its Generalizations and Applications}, 
    volume={1}, 
    DOI={10.1201/9781003168102}, 
    publisher={United Kingdom: Chapman & Hall}, author={Mezo, Istvan}, 
    year={2022} 
}

@article{Hutchings2017,
  title = {Tunable Superconducting Qubits with Flux-Independent Coherence},
  author = {Hutchings, M. D. and Hertzberg, J. B. and Liu, Y. and Bronn, N. T. and Keefe, G. A. and Brink, Markus and Chow, Jerry M. and Plourde, B. L. T.},
  journal = {Phys. Rev. Appl.},
  volume = {8},
  issue = {4},
  pages = {044003},
  numpages = {13},
  year = {2017},
  month = {10},
  publisher = {American Physical Society},
  doi = {10.1103/PhysRevApplied.8.044003},
  url = {https://link.aps.org/doi/10.1103/PhysRevApplied.8.044003}
}

@article{Leung2018,
  title = {Entangling an arbitrary pair of qubits in a long ion crystal},
  author = {Leung, Pak Hong and Brown, Kenneth R.},
  journal = {Phys. Rev. A},
  volume = {98},
  issue = {3},
  pages = {032318},
  numpages = {6},
  year = {2018},
  month = {09},
  publisher = {American Physical Society},
  doi = {10.1103/PhysRevA.98.032318},
  url = {https://link.aps.org/doi/10.1103/PhysRevA.98.032318}
}

@article{Babar2015,
  author={Babar, Zunaira and Botsinis, Panagiotis and Alanis, Dimitrios and Ng, Soon Xin and Hanzo, Lajos},
  journal={IEEE Access}, 
  title={Fifteen Years of Quantum LDPC Coding and Improved Decoding Strategies}, 
  year={2015},
  volume={3},
  number={},
  pages={2492-2519},
  doi={10.1109/ACCESS.2015.2503267}}

@article{breuckmann2021quantum,
  title={Quantum low-density parity-check codes},
  author={Breuckmann, Nikolas P and Eberhardt, Jens Niklas},
  journal={PRX Quantum},
  volume={2},
  number={4},
  pages={040101},
  year={2021},
  publisher={APS}
}

@article{bravyi2024high,
  title={High-threshold and low-overhead fault-tolerant quantum memory},
  author={Bravyi, Sergey and Cross, Andrew W and Gambetta, Jay M and Maslov, Dmitri and Rall, Patrick and Yoder, Theodore J},
  journal={Nature},
  volume={627},
  number={8005},
  pages={778--782},
  year={2024},
  publisher={Nature Publishing Group UK London}
}

@article{Liang2024,
   title={Pulse optimization for high-precision motional-mode characterization in trapped-ion quantum computers},
   volume={9},
   ISSN={2058-9565},
   url={http://dx.doi.org/10.1088/2058-9565/ad3a98},
   DOI={10.1088/2058-9565/ad3a98},
   number={3},
   journal={Quantum Science and Technology},
   publisher={IOP Publishing},
   author={Liang, Qiyao and Kang, Mingyu and Li, Ming and Nam, Yunseong},
   year={2024},
   month=apr, pages={035007} 
}

@article{Brown2011,
  title = {Single-qubit-gate error below ${\mathbf{10}}^{\ensuremath{-}\mathbf{4}}$ in a trapped ion},
  author = {Brown, K. R. and Wilson, A. C. and Colombe, Y. and Ospelkaus, C. and Meier, A. M. and Knill, E. and Leibfried, D. and Wineland, D. J.},
  journal = {Phys. Rev. A},
  volume = {84},
  issue = {3},
  pages = {030303},
  numpages = {4},
  year = {2011},
  month = {09},
  publisher = {American Physical Society},
  doi = {10.1103/PhysRevA.84.030303},
  url = {https://link.aps.org/doi/10.1103/PhysRevA.84.030303}
}

@article{Li2023,
  author       = {Zhiyuan Li and Pei Liu and Peng Zhao and Zhenyu Mi and Huikai Xu and Xuehui Liang and Tang Su and Weijie Sun and Guangming Xue and Jing-Ning Zhang and Weiyang Liu and Yirong Jin and Haifeng Yu},
  title        = {Error per single-qubit gate below 10$^{-4}$ in a superconducting qubit},
  journal      = {npj Quantum Information},
  year         = {2023},
  volume       = {9},
  number       = {1},
  pages        = {111},
  doi          = {10.1038/s41534-023-00781-x},
  url          = {https://doi.org/10.1038/s41534-023-00781-x}
}

@article{fowler2018low,
  title={Low overhead quantum computation using lattice surgery},
  author={Fowler, Austin G and Gidney, Craig},
  journal={arXiv preprint arXiv:1808.06709},
  year={2018}
}

@article{Fellous-Asiani2021,
  title = {Limitations in Quantum Computing from Resource Constraints},
  author = {Fellous-Asiani, Marco and Chai, Jing Hao and Whitney, Robert S. and Auff\`eves, Alexia and Ng, Hui Khoon},
  journal = {PRX Quantum},
  volume = {2},
  issue = {4},
  pages = {040335},
  numpages = {11},
  year = {2021},
  month = {11},
  publisher = {American Physical Society},
  doi = {10.1103/PRXQuantum.2.040335},
  url = {https://link.aps.org/doi/10.1103/PRXQuantum.2.040335}
}

@article{EFTQC_2021,
   title={Early fault-tolerant simulations of the Hubbard model},
   volume={7},
   ISSN={2058-9565},
   url={http://dx.doi.org/10.1088/2058-9565/ac3110},
   DOI={10.1088/2058-9565/ac3110},
   number={1},
   journal={Quantum Science and Technology},
   publisher={IOP Publishing},
   author={Campbell, Earl T},
   year={2021},
   month=nov, pages={015007} }

@article{EFTQC_2022,
  doi = {10.22331/qv-2022-07-22-65},
  url = {https://doi.org/10.22331/qv-2022-07-22-65},
  title = {Designing algorithms for estimating ground state properties on early fault-tolerant quantum computers},
  author = {Tong, Yu},
  journal = {{Quantum Views}},
  publisher = {{Verein zur F{\"{o}}rderung des Open Access Publizierens in den Quantenwissenschaften}},
  volume = {6},
  pages = {65},
  month = jul,
  year = {2022}
}

@article{EFTQC_2024,
  title = {Modeling the performance of early fault-tolerant quantum algorithms},
  author = {Liang, Qiyao and Zhou, Yiqing and Dalal, Archismita and Johnson, Peter},
  journal = {Phys. Rev. Res.},
  volume = {6},
  issue = {2},
  pages = {023118},
  numpages = {16},
  year = {2024},
  month = {05},
  publisher = {American Physical Society},
  doi = {10.1103/PhysRevResearch.6.023118},
  url = {https://link.aps.org/doi/10.1103/PhysRevResearch.6.023118}
}

@article{Q_catalysis_resources,
  title = {Quantum computing enhanced computational catalysis},
  author = {von Burg, Vera and Low, Guang Hao and H\"aner, Thomas and Steiger, Damian S. and Reiher, Markus and Roetteler, Martin and Troyer, Matthias},
  journal = {Phys. Rev. Res.},
  volume = {3},
  issue = {3},
  pages = {033055},
  numpages = {16},
  year = {2021},
  month = {Jul},
  publisher = {American Physical Society},
  doi = {10.1103/PhysRevResearch.3.033055},
  url = {https://link.aps.org/doi/10.1103/PhysRevResearch.3.033055}
}

@article{lee_even_2021,
        title = {Even {More} {Efficient} {Quantum} {Computations} of {Chemistry} {Through} {Tensor} {Hypercontraction}},
        volume = {2},
        url = {https://link.aps.org/doi/10.1103/PRXQuantum.2.030305},
        doi = {10.1103/PRXQuantum.2.030305},
        number = {3},
        journal = {PRX Quantum},
        author = {Lee, Joonho and Berry, Dominic W. and Gidney, Craig and Huggins, William J. and McClean, Jarrod R. and Wiebe, Nathan and Babbush, Ryan},
        year = {2021},
        note = {Publisher: American Physical Society},
        pages = {030305},
}

@article{cleve1998quantum,
  title={Quantum algorithms revisited},
  author={Cleve, Richard and Ekert, Artur and Macchiavello, Chiara and Mosca, Michele},
  journal={Proceedings of the Royal Society of London. Series A: Mathematical, Physical and Engineering Sciences},
  volume={454},
  number={1969},
  pages={339--354},
  year={1998},
  publisher={The Royal Society}
}

\end{document}